\documentclass[aps,prl,twocolumn,showpacs,lengthcheck]{revtex4}%
\usepackage{amsfonts}
\usepackage{amsmath}
\usepackage{amssymb}
\usepackage{graphicx}%
\begin{document}
\title{Dynamical evolution of a doubly-quantized vortex imprinted in a Bose-Einstein Condensate}
\author{A. Mu\~{n}oz Mateo}
\email{ammateo@ull.es}
\author{V. Delgado}
\email{vdelgado@ull.es}
\affiliation{Departamento de F\'{\i}sica Fundamental II, Universidad de La Laguna}

\pacs{03.75.Kk, 03.75.Lm, 67.90.+z}

\begin{abstract}
The recent experiment by Y. Shin \emph{et al.} [Phys. Rev. Lett. \textbf{93},
160406 (2004)] on the decay of a doubly quantized vortex imprinted in $^{23}%
$Na condensates is analyzed by numerically solving the Gross-Pitaevskii
equation. Our results, which are in very good quantitative agreement with the
experiment, demonstrate that the vortex decay is mainly a consequence of
dynamical instability. Despite apparent contradictions, the local density
approach is consistent with the experimental results. The monotonic increase
observed in the vortex lifetimes is a consequence of the fact that, for large
condensates, the measured lifetimes incorporate the time it takes for the
initial perturbation to reach the central slice. When considered locally, the
splitting occurs approximately at the same time in every condensate,
regardless of its size.

\end{abstract}
\date{3 June 2006}
\maketitle


Since the creation of the first dilute-gas Bose-Einstein condensates there has
been great interest in characterizing their superfluid properties. This has
stimulated a great deal of theoretical and experimental work aimed at studying
the rotational properties of dilute Bose gases \cite{Fetter1} and, in
particular, the nucleation and stability properties of vortices \cite{SV1}.
Numerous experiments have succeeded in generating vortices \cite{Vortices}. In
practically all of them, the vorticity appears concentrated in a number of
singly quantized vortices. This is a consequence of the fact that multiply
quantized vortices are energetically unstable against their splitting in an
array of vortices with unit topological charge \cite{Rokh1}. Multiply
quantized vortices are also dynamically unstable, which implies that they
decay even in the zero-temperature limit \cite{Pu1}.

Recently, Leanhardt \emph{et al.} \cite{Lean1} obtained multiply quantized
vortices in a gaseous Bose-Einstein condensate by using a topological
phase-imprinting technique proposed by Nakahara \emph{et al.} \cite{Nak1}.
They started from nonrotating $^{23}$Na condensates in the $|1,-1>$ and
$|2,+2>$ hyperfine states. By adiabatically inverting the magnetic bias field,
the initial states were transformed into vortex states with axial angular
momentum per particle $2\hbar$ and $-4\hbar$, respectively. This has opened
the possibility for studying\ experimentally the stability of multiply
quantized vortices, and has stimulated new theoretical works. In particular,
M\"{o}tt\"{o}nen \emph{et al. }\cite{Moto1}\ have studied numerically the
stability of a doubly quantized vortex in a cylindrical condensate as a
function of the (dimensionless) interaction strength per unit length along the
condensate axis, $an_{z}$, where $a$ is the s-wave scattering length and
$n_{z}=%
{\textstyle\int}
\left\vert \Psi(\mathbf{r)}\right\vert ^{2}dx\,dy$ is the linear atom density
along $z$. They found a series of quasiperiodic instability regions in the
parameter space of $an_{z}$ \cite{Pu1}. The first two of them (the only ones
relevant to this work) correspond to $an_{z}<3$ and $11.4\lesssim
an_{z}\lesssim16$, respectively. By comparing with the solution of the
Gross-Pitaevskii equation for a harmonically trapped cigar-shaped condensate
these authors conclude that a doubly quantized vortex is dynamically unstable,
and it is in the above instability regions where the vortex decay is
initiated, giving rise in the process to a pair of intertwining
singly-quantized vortices.

Shortly after, a second experiment was carried out at MIT \cite{Shin1}, aimed
at studying the characteristic time scale of the splitting process as a
function of the interaction strength $an_{z=0}$. Experimental results were
obtained from integrated absorption images along the condensate axis after
$15$ $%
\operatorname{ms}%
$ of ballistic expansion. A key ingredient was the fact that in order to
increase the visibility of the vortex cores, absorption images were restricted
to a $30$ $%
\operatorname{\mu m}%
$ thick central slice of the condensate. This experiment showed that, even at
$T\simeq0$, doubly quantized vortices decay into two singly quantized vortices
on a time scale that is longer at higher atom density, with an observed
lifetime that increases monotonically with $an_{z=0}$, showing no
quasiperiodic behavior.

These results, pose a number of open questions. First, the observed monotonic
behavior seems to be not consistent with the local density approach proposed
in Ref \cite{Moto1}. According to this proposal, as the atom density
increases, for $an_{z=0}\sim12$, the second instability region mentioned above
arises within the central slice. One then would expect this fact to manifest
as a decrease in the observed lifetime, which does not occur. This raises some
doubts on the validity of the local density approach \cite{Shin1}. Second,
numerical work has appeared recently suggesting that the vortex decay is a
consequence of thermal fluctuations \cite{Gawr1}. Even though it seems
reasonable to attribute the splitting to a dynamical instability, it would be
desirable to have theoretical results in good quantitative agreement with the
experiment, which could corroborate this assumption. Finally, the detailed
dynamical behavior of the vortex along the entire $z$-axis can be especially
relevant for characterizing the splitting process in very elongated
condensates such as those studied in the MIT experiment. Different regions
along the $z$-axis could exhibit different local behavior which could be
determinant to understand the process. The experimental results cannot provide
this kind of information, and thus it would be desirable to complement them
with the detailed dynamics of the vortex along the entire condensate.

In this Letter we address the above questions by performing a realistic
computer simulation of the MIT experiment. Our results, which are in very good
quantitative agreement with the experiment, enable us to understand the
experimental results by providing a detailed visualization of the splitting
process. They confirm that the decay of the vortex core is essentially a
dissipationless process driven by a dynamical instability. They also indicate
that the local density picture proposed in Ref. \cite{Moto1} provides the key
ingredients to interpret properly the splitting process. When considered
locally, the decay is characterized by the sole parameter $an_{z}$ and it is
always initiated in those regions along the $z$-axis where $an_{z}\simeq1.5$
or $an_{z}\simeq13.75$, on a time scale that is in all cases of the order of
$15$ $%
\operatorname{ms}%
$. For large condensates ($an_{z=0}\gtrsim3$), the lifetime observed in the
central slice incorporates the time it takes for the nonlinear perturbation to
propagate from the location where the splitting is initiated to the final
position where it is eventually detected. This nonlocal process explains the
observed monotonic increase in the vortex lifetime and makes the local density
picture compatible with the experimental results.

In the zero-temperature limit, the dynamics of dilute Bose gases is accurately
described by the Gross-Pitaevskii equation (GPE), which governs the time
evolution of the condensate wave function%

\begin{equation}
i\hbar\frac{\partial\Psi}{\partial t}=\left(  -\frac{\hbar^{2}}{2m}\nabla
^{2}+V(\mathbf{r})+gN\left\vert \Psi(\mathbf{r},t)\right\vert ^{2}\right)
\Psi\label{eq1}%
\end{equation}
where $N$ is the number of atoms, $g=4\pi\hbar^{2}a/m$ is the interaction
strength, and $V(\mathbf{r})=\frac{1}{2}m(\omega_{r}^{2}+\omega_{z}^{2})$ is
the external confining potential.

For the $^{23}$Na condensates used in the MIT experiment, $a=2.75$ $%
\operatorname{nm}%
$ and $\omega_{r}/2\pi=220$ $%
\operatorname{Hz}%
$. The experimental results were obtained using, without distinction, three
different axial trap frequencies, $\omega_{z}/2\pi=2.7$, $3.7$, and $12.1$ $%
\operatorname{Hz}%
$. Integration of the 3D GPE for such highly elongated condensates is a very
demanding computational task. As the system evolves in time it develops a
complex fine structure that can only be properly resolved by using very large
basis or gridpoint sets. To verify the convergence and accuracy of our
numerical results we have implemented two different integration methods: a
Laguerre-Fourier and a Laguerre-Hermite-Fourier pseudospectral method. The
evolution in time has been carried out by a third-order Adams-Bashforth
time-marching scheme. The same results have been obtained with the two
integration methods and using very different basis sets and time steps.

\begin{figure}
[ptb]
\begin{center}
\includegraphics[
height=5.1796cm,
width=7.295cm
]%
{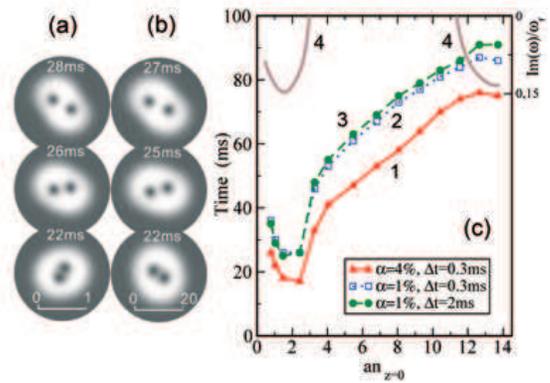}%
\caption{(color online) (a) In-trap absorption images of a condensate with
$an_{z=0}=1.48$. (b) Same as (a) after $15$ $\operatorname{ms}$ of ballistic
expansion. Lengths are in units of the axial trap, $a_{z}=6.05$
$\operatorname{\mu m}$. (c) Predicted splitting times as a function of
$an_{z=0}$ for three different perturbations (curves 1-3). Curve 4 shows the
excitation spectrum of the unstable modes.}%
\label{Fig1}%
\end{center}
\end{figure}

We have solved the GPE starting, in all cases, from the stationary state
compatible with a doubly quantized vortex. A quantum system is dynamically
unstable when it is unstable against arbitrarily small perturbations. To
determine whether the vortex decay can be unambiguously attributed to a
dynamical instability, at $t=0$, we introduce a small fluctuation in the
trapping potential. Specifically, we introduce a $1\%$ quadrupolar
perturbation for a short interval of $0.3%
\operatorname{ms}%
$. Such a small perturbation produces an almost undetectable change in both
the energy and\ the angular momentum per particle ($\Delta E/E,\,\Delta
L_{z}/L_{z}<10^{-5}$). In order to compare with the experiment we have
followed the same procedure as used at MIT. Condensates with different values
of $an_{z=0}$ were produced by varying $N$, and integrated absorption images
of a $30$ $%
\operatorname{\mu m}%
$ thick central slice were then generated at different times. In all cases the
initial vortex decayed into a pair of singly quantized vortices. We have also
considered the potential effect of dissipative processes by introducing a
phenomenological imaginary time and, for values consistent with the
experiment, we have found it to be negligible. The vortex lifetime was
inferred from the absorption images by identifying the instant at which two
vortex cores can be resolved unambiguously. An example is shown in Fig.
\ref{Fig1}a. We have also analyzed the effect of\ the ballistic expansion by
solving numerically the corresponding GPE in a number of representative cases.
This process modifies the predicted lifetime by only $1-2%
\operatorname{ms}%
$ approximately (Fig. \ref{Fig1}b), which is a consequence of the fact that
most of the expansion is a mere scale transformation. Since this correction is
of the order of measurement uncertainties we neglect it in what follows.%

To investigate the effects of both\ the strength and the duration of the
perturbation we have considered two additional cases: a $4\%$ and a $1\%$
quadrupolar perturbation acting for $0.3$ $%
\operatorname{ms}%
$ and $2$ $%
\operatorname{ms}%
$, respectively. The corresponding numerical results, obtained for an axial
trap frequency $\nu_{z}=12$ $%
\operatorname{Hz}%
$, are shown in Fig. \ref{Fig1}c. Even though the local value of $an_{z}$
essentially determines where and when the splitting starts, the sole parameter
$an_{z=0}$ is not sufficient to characterize completely the observed splitting
times in the central slice. For instance, for the $4\%$ quadrupolar
perturbation the predicted splitting time for a condensate with $an_{z=0}%
=3.27$ in the trap with $\nu_{z}=3.7$ $%
\operatorname{Hz}%
$ turns out to be $55$ $%
\operatorname{ms}%
$, whereas for a condensate with the same value of $an_{z=0}$\ but in the $12$
$%
\operatorname{Hz}%
$ trap one obtains $33$ $%
\operatorname{ms}%
$. In general, for $an_{z=0}\gtrsim3$, the more elongated the condensate, the
longer the measured lifetime. This is a consequence of the fact that measured
lifetimes incorporate the time it takes for the initial perturbation to reach
the central slice. Since the shortest lifetimes occur for the largest axial
frequency ($12$ $%
\operatorname{Hz}%
$), this is the only case one has to consider: experimental data corresponding
to two visible cores must lie above the theoretical curve. Note that this does
not prevent data corresponding to a single core (those obtained with $\nu
_{z}=2.7$ or $3.7$ $%
\operatorname{Hz}%
$) from also lying above the theoretical curve.

\begin{figure}
[ptb]
\begin{center}
\includegraphics[
height=5.1796cm,
width=7.295cm
]%
{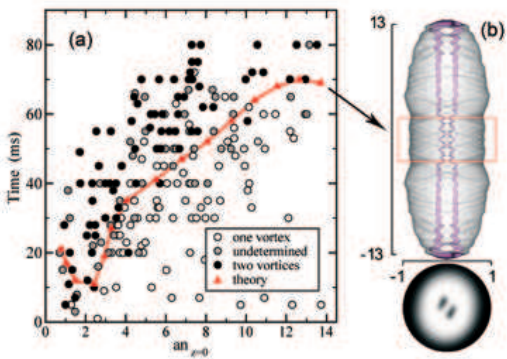}%
\caption{(color online) (a) Predicted splitting times as a function of
$an_{z=0}$ for a $4\%$ quadrupolar perturbation acting during $0.3$
$\operatorname{ms}$. (b) Density isosurface of a condensate with
$an_{z=0}=13.75$ at $t=75$ $\operatorname{ms}$. The corresponding axial
absorption image of the central slice (rectangle in the figure) is also shown.
Lengths are in units of $a_{z}=6.05$ $\operatorname{\mu m}$.}%
\label{Fig2}%
\end{center}
\end{figure}

While the perturbation strength has an important influence on the predicted
lifetimes, its duration seems to be of little importance (Fig. \ref{Fig1}c).
The main effect of stronger perturbations is to shift the predicted curve
toward shorter times. The best agreement with the experiment is found for the
$4\%$ quadrupolar perturbation (curve 1). However, the predicted lifetimes are
somewhat longer than the experimental ones. Note that the experimental results
do not include the $12%
\operatorname{ms}%
$ spent in the inversion of the axial magnetic field $B_{z}$ (the vortex
imprinting). Since the vortex already forms as $B_{z}=0$ \cite{Ogawa1}, from
this process (the preparation time) one reasonably can expect a contribution
of about $5-6%
\operatorname{ms}%
$. Fig. \ref{Fig2} shows the corresponding theoretical curve (from which we
have subtracted a conservative amount of $6%
\operatorname{ms}%
$ to account for the preparation time) along with the experimental data of the
MIT experiment. The good agreement demonstrates that the decay is mainly
driven by a dynamical instability.%

During the vortex preparation, as $B_{z}$ vanishes, all of the three $F=1$
components appear in the trap. Thus, for a short interval around the instant
of preparation the spinorial character of the multicomponent BEC cannot be
neglected \cite{Ogawa1}. The weak-field seeking state becomes necessarily
perturbed by the other components. In the presence of gravitational
interaction (perpendicular to the $z$-axis in the experimental setup) each
component is shifted from the $z$-axis by a different amount \cite{Kawa1}.
This introduces an effective non-axially symmetric perturbation on the
dynamical evolution of the weak-field seeking component via the corresponding
interaction terms. It seems reasonable to think that the quadrupolar component
of this complex effective perturbation might be responsible for the vortex decay.

The theoretical curve in Fig. \ref{Fig2} displays a clear minimum about
$an_{z=0} \simeq1.5$. It is instructive to put this minimum in relation to the
excitation spectrum of the unstable (imaginary frequency) modes (see curve 4
in Fig. \ref{Fig1}c) \cite{Moto1,Pu1}. For small condensates with
$an_{z=0}<1.5$ all of the $z$-axis lies within the first instability region.
For such condensates only a limited subset of the unstable modes associated
with that instability region become accessible. The smaller the condensate the
smaller the maximum unstable frequency and, consequently, the longer the
observed decay time. For condensates with $an_{z=0}\geq1.5$ all of the
unstable frequencies become accessible. In such condensates, as the number of
particles increases the first instability region (corresponding to those
values along the $z$-axis where $an_{z}<3$) moves progressively away from the
central slice, towards the edges of the condensate, occupying symmetric
positions around $z=0$. As a consequence, the splitting of the vortex core has
to propagate from those regions where it is initiated to the central slice
before it can be detected. This nonlocal process is responsible for the
monotonic increase in the lifetimes for $an_{z=0}\gtrsim3$, and explains the
minimum predicted about $an_{z=0}\simeq1.5$. According to a local density
picture, in principle, one also would expect a second minimum about
$an_{z=0}\simeq13.75$. Even though no such a minimum occurs, a clear change in
the slope of the predicted curve can be identify as $an_{z=0}$ enters the
second instability region, indicating a different physical behavior.

\begin{figure}
[ptb]
\begin{center}
\includegraphics[
height=3.7255cm,
width=5.5926cm
]%
{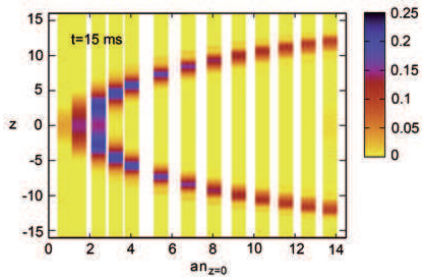}%
\caption{(color online) Density along the $z$-axis as a function of $an_{z=0}$
at $t=15\operatorname{ms}$. Lengths are in units of $a_{z}=6.05$
$\operatorname{\mu m}$.}%
\label{Fig3}%
\end{center}
\end{figure}

In Fig. \ref{Fig3} we have calculated the density along the $z$-axis of every
condensate (characterized by the different $an_{z=0}$), at the same instant of
time ($t=15%
\operatorname{ms}%
$). This axial density has been renormalized in each $z$-plane\ in such a way
that the maximum density in that plane is $1$. Thus, for a given $an_{z=0}$,
dark zones in the plot density indicate those regions along the condensate
axis where the vortex splitting has already begun. As is evident from Fig.
\ref{Fig3}, when considered as a local process, the splitting takes place
approximately at the same time (on a timescale of the order of $15%
\operatorname{ms}%
$) in every condensate, regardless of its size. In all cases the process
starts precisely at the predicted instability regions. Fig. \ref{Fig3} also
shows that for small condensates with $an_{z=0}<3$ the splitting occurs almost
simultaneously along the entire $z$-axis, which is a consequence of the fact
that in such condensates the entire $z$-axis lies within the first instability
region. As a result, the initial vortex decays into a pair of nearly straight
unit-charge vortices. Note that, for $an_{z=0}<1.5$ the splitting takes
somewhat longer times, as expected. As the condensate size increases, the
vortex splitting becomes a localized phenomenon that propagates along the
$z$-axis, producing in this case a pair of intertwining singly-quantized
vortices \cite{Moto1}, which is a consequence of the dephase in time between
different $z$-planes. For $an_{z=0}>12$ a new instability region appears at
the center of the condensate and the vortex splitting is firstly initiated at
the edges of the condensate and a few milliseconds later at the central slice
(a consequence of the smaller value of the corresponding maximum unstable frequency).%

\begin{figure}
[ptb]
\begin{center}
\includegraphics[
height=6.0056cm,
width=5.6936cm
]%
{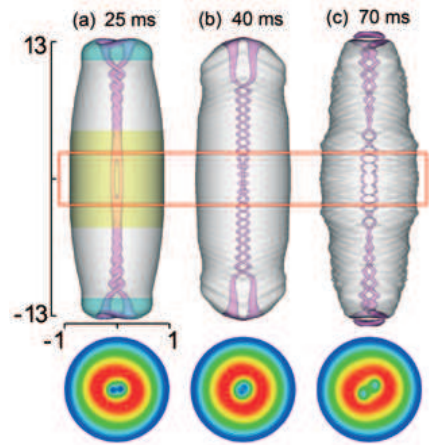}%
\caption{(color online) Evolution in time of the splitting process of a
condensate with $an_{z=0}=13.75$. The shaded zones in (a) indicate the
instability regions. The corresponding axial absorption images of the central
slice (rectangle in the figure) are also shown. Lengths are in units of
$a_{z}=6.05$ $\operatorname{\mu m}$.}%
\label{Fig4}%
\end{center}
\end{figure}

Fig. \ref{Fig4} shows the evolution in time of the splitting process for a
condensate with $an_{z=0}=13.75$. The first and second instability regions
correspond to the shaded zones in Fig. \ref{Fig4}a. This figure clearly shows
that, at $t=25%
\operatorname{ms}%
$, the splitting process has already begun in both the edges and the center of
the condensate. However, the two vortex cores overlap (Fig. \ref{Fig4}b) and
thus cannot be experimentally resolved until much longer times. At $t\approx70%
\operatorname{ms}%
$ (Fig. \ref{Fig4}c) the vortex cores begin to disentangle in such a way that
they can be unambiguously resolved at $t=75%
\operatorname{ms}%
$ (Fig. \ref{Fig3}b). Thus, despite appearances no contradiction exists with
the local density picture. No decrease is observed in the decay times due to
the fact that the time required to resolve the two vortex cores is
much\ longer than that it takes for the instability originating at the edges
of the condensate to reach the central slice.%

In conclusion, we have analyzed the MIT experiment by solving the full GPE.
Our results, which are in very good quantitative agreement with the
experiment, demonstrate that the vortex decay is mainly driven by a dynamical
instability, and allow a better understanding of the splitting process.
Despite apparent contradictions, the local density approach is consistent with
the experimental results. The monotonic increase observed in the vortex
lifetimes is a consequence of the fact that, for large condensates, the
measured lifetimes incorporate the time it takes for the initial perturbation
to reach the central slice. When considered locally, the splitting occurs
approximately at the same time in every condensate, regardless of its size.

\begin{acknowledgments}
This work has been supported by Ministerio de Educaci\'{o}n y Ciencia (Spain)
and FEDER fund (EU) (contract No. Fis2005-02886).
\end{acknowledgments}

Note added--After this work was finished, we learned about a very recent
preprint [quant-ph/0605125] in which the same experiment is analyzed.

\bigskip

\end{document}